\documentclass[9pt,showpacs]{revtex4}

\usepackage{epsfig}
\input epsf.tex
\usepackage{amssymb}
\usepackage{amsmath}
\usepackage{amsthm}
\usepackage{amsopn}

\def\comment#1{}

\def\beq{\begin{equation}}
\def\eeq{\end{equation}}
\def\bea{\begin{eqnarray}}
\def\eea{\end{eqnarray}}

\usepackage{epsfig}
\input epsf.tex
\usepackage{amssymb}
\usepackage{amsmath}
\usepackage{amsthm}
\usepackage{amsopn}

\def\comment#1{}

\def\beq{\begin{equation}}
\def\eeq{\end{equation}}
\def\bea{\begin{eqnarray}}
\def\eea{\end{eqnarray}}

\usepackage{ulem}
\usepackage{cancel}
\usepackage{color}

\usepackage{ulem}
\usepackage{cancel}
\usepackage{color}

\begin{document}

\title{ Generation of circular polarization in CMB radiation via nonlinear photon-photon interaction}
\author{Mehdi Sadegh $^1$}
\email{m.sadegh-AT-pnu.ac.ir}
\author{Rohoollah Mohammadi$^{2,3}$}
\email{rmohammadi-AT-ipm.ir}

\author{ Iman Motie$^{4}$}

\affiliation{$^1$Department of Physics, Payame Noor University (PNU), P.O. Box 19359-3697, Tehran, Iran},

\affiliation{$^2$Iranian National Science and Technology Museum (INMOST), PO BOX: 11369-14611, Tehran, Iran,}

\affiliation{$^3$ School of Astronomy, Institute for Research in Fundamental
Sciences (IPM), P. O. Box 19395-5531, Tehran, Iran,}

\affiliation{$^4$ Department of Physics, Mashhad Branch, Islamic Azad University, Mashhad, Iran.}

\date{\today}% 1em {\it  Department  of Physics, Mashhad Branch, Islamic Azad University, Mashhad, Iran }

\begin{abstract}
 Standard  cosmological  models  do  predict  a  measurable  amount  of anisotropies  in  the  intensity  and  linear  polarization  of
the  Cosmic  Microwave  Background radiation (CMB) via Thomson scattering, even though these theoretical models do not  predict circular  polarization for CMB radiation. In other hand, the circular polarization of CMB has not been excluded in observational evidences. Here we estimate the circular polarization power spectrum $C_{l}^{V(S)}$ in CMB radiation due to Compton scattering and non-linear photon-photon forward scattering via Euler-Heisenberg Effective Lagrangian. We
have estimated the average  value of circular power spectrum is $1(l+1)\,C_{l}^{V(S)}/(2\pi)\sim 10^{-4}\mu\,K^2$ for $l\sim300$ at present time which is smaller than recently reported data (SPIDER collaboration) but in the range of the future achievable experimental data. We also show that the generation of B-mode polarization for CMB photons in the presence of the primordial scalar perturbation via Euler-Heisenberg interaction is possible however this contribution for B-mode polarization is not remarkable.

\end{abstract}
\maketitle
%%%%%%%%%%%%%%%%%%%%%%%%%%%%%%%%%%%%%%%%%%%%%%%%%%%%%%%%%%%%%%%%%%%%%%%%%%%%%%%%%%%%%%%%%%%%%%%%%%%%%%%%%%%%%%%%%%%%%%%%%%%
\newpage
\section{\large Introduction}
 Photon-matter interactions can convert or generate the polarization states of photons in different situation such as Faraday rotation, Faraday conversion and so on.
In some special cases, the measurement of circular polarization contribution provides very important tools to better understand universe.
In standard scenario of cosmology, CMB anisotropies are partially linearly polarized \cite{nature,exp1,num1,num2,dasi,cosowsky1994} while the generation of circular polarization is ignored, because there is no a notable mechanism to generate circular polarization in the recombination epoch.
Note  Compton (Thomson) scattering, as most important interaction of CMB radiation, cannot
generate the circular polarization \cite{cosowsky1994}.\\
In other hand, the circular polarization of CMB has not been excluded in observational evidences. For example, recently SPIDER collaboration are
made maps of approximately $10\%$ of the sky with degree-scale angular resolution in 95 and
150 GHz observing bands.  Data of SPIDER group have been  analyzed in \cite{spider} and a new upper limit
on CMB circular polarization is obtained, so that  constrains of the circular power spectrum $l(l+1)\,C_l^V/(2\pi)$ are reported in rang of  a few hundred $\mu K^2$ at 150 GHz for a thermal CMB spectrum. Also it is worthwhile take a look  other  reports about
the constraint  on  the circular polarizations $\frac{\Delta V}{T_{CMB}}$ \cite{exp0,exp2,exp3} and B-mode polarization \cite{exp4,exp5,exp6}.\\
In the case of theoretical models,  there
are several mechanisms, almost considering new physics interactions, which discuss the possibility of the generation of
circular polarization in the CMB. For instances, the conversion of the existing  linear  polarization  into  circular  one  in the presence of external magnetic fields of galaxy clusters \cite{fc}, the relativistic plasma remnants \cite{tashiro} and   magnetic  fields  in  the  primordial
universe \cite{khodam,gio1,gio2} is discussed.  Forward scattering of CMB radiation
from the cosmic neutrino background \cite{mohammadi},
and photon-photon interactions in neutral hydrogen \cite{sawyer} have also been shown as potentially mechanisms for the generation of CMB circular polarization. There are some mechanisms which are postulated extensions to QED such as Lorentz-invariance violating operators \cite{khodam,colladay, lv cos},
axion-like pseudoscalar particles \cite{finelli},
and non-linear photons interactions (through effective Euler-
Heisenberg Lagrangian) \cite{ix}. In \cite{sims}, the production of primordial circular polarization in axion inflation coupled
to fermions and gauge fields, with special attention paid to reheating, have been studied. Also see a brief review  of some of the mentioned mechanisms in \cite{king}.

In this work, we focus on the generation of circular polarization due to nonlinear photon-photon interaction (via Euler-Heisenberg Lagrangian). Of course we should mention that Faraday conversion phase shift $\Delta\phi_{FC}$ due to Euler-Heisenberg Lagrangian for CMB radiation has been estimated in \cite{ix}.  It is worthwhile to mention that one can calculate $\Delta\phi_{FC}$ from below equation [see more detail in \cite{fc,faraday con}]
\begin{eqnarray}
\dot{V}=2U\frac{d}{dt}(\Delta\phi_{FC}),
\label{fc}
\end{eqnarray}
where $U$ and $V$ are Stokes parameters which describe linear and circular polarizations respectively. Note $\Delta\phi_{FC}$ reported in \cite{ix} is not suitable quantity to compare with experimental data which usually reported by circular polarization power spectrum $C_l^V$. So the main purpose of our work is to calculate $C_l^V$ via Euler-Heisenberg effective interactions and make a comparison with recently data reported by SPIDER collaboration group.

We start by a brief discussion on Stokes parameters and their definitions in terms of density matrix elements. Then we calculate time evolution of those parameters by Euler-Heisenberg consideration. In the next two sections we solve them by some estimations to calculate dominant contribution terms. This contributions come from total intensity of CMB photon contribution in comparison with linear and circular polarizations. Finally in the last section, we compute the power spectrum and B-mode spectrum of CMB photons which are generated by Euler-Heisenberg effective Lagrangian.

\section{\large Polarization and Stokes parameters}
An ensemble of photons in a completely general mixed states
is given by a normalized density matrix $\rho_{ij}\equiv (\,|\varepsilon_i\rangle\langle \varepsilon_j|/{\rm tr}\rho)$,
where in the quantum mechanics description, an arbitrary polarized
state of a photon with energy $(|k^0|^2=|{\bf k}|^2)$ propagating in the $\hat
z$-direction is written as
\begin{eqnarray}
|\varepsilon\rangle=a_1\exp(i\theta_1)|\varepsilon_1\rangle+a_2\exp(i\theta_2)|\varepsilon_2\rangle,
\end{eqnarray}
where $|\varepsilon_1\rangle$ and $|\varepsilon_2\rangle$ represent the polarization states in the $\hat{x}$-
and $\hat{y}$-directions. Then
 the $2\times 2$ density matrix $\rho$ of photon polarization states are given as
\begin{eqnarray}\label{stokesparameter}
\rho=\left(\begin{array}{cc}
          \rho_{11} & \rho_{12} \\
          \rho_{21} & \rho_{22}\\
          \end{array}\right)=
\frac{1}{2}\left(
                   \begin{array}{cc}
                     I+Q & U-iV \\
                     U+iV & I-Q \\
                   \end{array}
                 \right),
\end{eqnarray}
where
$I, Q, U$ and $V$ are Stokes parameters, so that  $I$-parameter is the
total intensity of radiation, $Q$- and $U$-parameters indicate the intensity of linear
polarization of radiation, and $V$-parameter
determines the intensity of circular polarization of radiation. Note  $I$ and $V$ are independently physical observable quantities of the coordinate system, while $Q$- and $U$-parameters depend on the orientation of the selected coordinate system. Linear polarization can also be
characterized through a vector parameter $\mathbf{P}$ which describe by $|\mathbf{P}|\equiv\sqrt{Q^2+U^2}$ and $\alpha=\frac{1}{2}\tan^{-1}\frac{U}{Q}$ \cite{jackson}.\\
The time evolution of each Stokes parameter can be yielded through the
Quantum Boltzmann equation. To do this issue, ones can paly with each polarization state of
the CMB radiation as the phase space distribution function $\chi$ which can generally obey from
the classical Boltzmann equation
\begin{equation}\label{B-E}
    \frac{d}{dt}\chi=\mathcal{C}(\chi).
\end{equation}
The left hand side of above equation is known as the Liouville term (containing all gravitational effects), while the right hand side one contains all possible collision terms. By considering the CMB interactions on the right hand side of Boltzmann equation, we can calculate the time evolution of  the each polarization state of the photons. In the next section, we consider non-linear photon-photon forward
scattering via the Euler-Hesinberg Hamiltonian to compute the time evolution of each polarization sates.
%%%%%%%%%%%%%%%%%%%%%%%%%%%%%%%%%%%%%%%%%%%%%%%%%%%%%%%%%%%%%%%%%%%%%%%%%%%%%%%%%%%%%%%%%%%%%%%%%%%%%%%%%%%%

\section{\large The Euler-Heisenberg Lagrangian and The Photons Polarizations}
 The time evolution of $\rho_{ij}(k)$s as well
as Stokes parameters are given by [see \cite{cosowsky1994} for more detail],
\begin{eqnarray}
(2\pi)^3 \delta^3(0)(2k^0)\frac{d}{dt}\rho_{ij}(k) = i\langle
\left[H^0_I (t);D^0_{ij}(k)\right]\rangle-\frac{1}{2}\int
dt\langle \left[H^0_I(t);\left[H^0_I
(0);D^0_{ij}(k)\right]\right]\label{bo}\rangle,
\end{eqnarray}
where $H^0_I(t)$ is the leading order of the photon-photon interacting via
Euler-Hiesenberg Hamiltonian. The first term on the right-handed side of above equation is called forward
scattering term, and the second one is a higher order collision
term which is in order of the ordinary  cross section of photon-photon scattering.
 The Euler-Heisenberg Lagrangian is a low energy
effective lagrangian describing multiple photon interactions.
The first order of  photon-photon interacting hamiltonian via  Euler-Heisenberg Lagrangian can be written as \cite{euler,wisk}
\begin{eqnarray}
H^0_I(t) &=& -\frac{\alpha^2}{90m^4}\int d^3x\Big[
(F_{\mu\nu}{F}^{\mu\nu} )^2+\frac{7}{4}(F_{\mu\nu}\tilde{F}^{\mu\nu})^2\Big],
\label{int}
\end{eqnarray}
where $F_{\mu\nu}=\partial_\mu A_\nu-\partial_\nu A_\mu$ is the strength of
electromagnetic field and
$\tilde{F}^{\mu\nu}=\epsilon^{\mu\nu\alpha\beta}F_{\alpha\beta}$,
in which $\epsilon^{\mu\nu\alpha\beta}$ is an antisymmetric tensor
of rank four [for example see \cite{xue} and \cite{dunne}]. Note
\begin{align}\label{A00}
\hat A_{\mu}(x)=\int\frac{d^{3}\mathbf{k}}{(2\pi)^3 2k^{0}}[a_{r}(k)\epsilon_{r\mu}
(k)e^{-ik.x}+a_{r}^{\dagger}(k)\epsilon_{r\mu}^{\ast}(k)e^{ik.x}].
\end{align}
where creation $a_{r}^{\dagger}$ and annihilation $a_{r}$ operators satisfy the canonical commutation relation as
\begin{align}\label{e22}
[a_{r}(k) , a_{r'}^{\dagger}(k^{'})]=(2\pi)^{3}2k^{0}\delta_{rr'}\delta^{(3)}(\mathbf{k}-\mathbf{k}^{'}).
\end{align}
We only compute the first order of Quantum Boltzmann Equation i.e.
the first term in RHS of the Eq.~(\ref{bo}), and neglect the
second term which is in order of $\alpha^4$. In principle when
first term doesn't have any result, in any special theory, one can try
to compute the second term.
It is worthwhile to mention that the contribution of $(F_{\mu\nu}\tilde{F}^{\mu\nu})^2$ for CMB polarization is given in \cite{ix}, however they have just calculated Faraday Conversion phase shift. Here we will consider both term of Euler-Heisenberg Lagrangian.
After tedious but straightforward calculation, using
Eq.~(\ref{j1}), the time-evolutions of Stokes
parameters Eq.(\ref{stokesparameter}) are obtained (find details in Appendix). First we start with $I$-parameter
\begin{eqnarray}
\dot{I}(\mathbf{k})%&=&\dot\rho_{11}+\dot\rho_{22}
&=&0,\label{id}
\end{eqnarray}
 $\dot{I}(\mathbf{k})=0$ implies, for each ensemble of photons like CMB,
the total intensity $I$ in any direction $\hat{\mathbf{k}}$ is constant and does not change from Euler-Heisenberg forward scattering. The above result for intensity $I$ is expected, because the forward scattering  cannot change momenta of photons which is necessary condition to change intensity in any direction. Note for the rest of paper, we do not consider the terms with linearly dependence of $\rho_{ij}$ on the right side of above equations, because we are interested in photon-photon forward scattering. The time evolution of linear and circular polarization parameters are given as following
\begin{eqnarray}
\dot{Q}(\mathbf{k})%&=&\dot\rho_{11}-\dot\rho_{22}\nonumber\\
&=&\frac{16\alpha^2}{45m^4 k^0}V(\mathbf{k})\!\!\!\int\!\!\!\frac{d^3p}{(2\pi)^3 2p^0}(p^0k^0)^2
\Big[f_1(\hat{p},\hat{k})U(\mathbf{p})\Big],
\label{qd}
\end{eqnarray}
\begin{eqnarray}
\dot{U}(\mathbf{k})%&=&\dot\rho_{12}+\dot\rho_{21}\nonumber\\
&=&\frac{8\alpha^2}{45m^4 k^0}V(\mathbf{k})\!\!\!
\int\frac{d^3p}{(2\pi)^32p^0}(p^0k^0)^2\Big[
f_1(\hat{p},\hat{k})I(\mathbf{p})\Big].
\label{ud}
\end{eqnarray}
\begin{eqnarray}
\dot{V}(\mathbf{k})%&=&i(\dot\rho_{21}-\dot\rho_{12})\nonumber\\
&=&\frac{8\alpha^2}{45m^4 k^0}U(\mathbf{k})\!\!\! \int\frac{d^3p}{(2\pi)^32p^0}(p^0k^0)^2
\Big[f_2(\hat{p},\hat{k})I(\mathbf{p})
\Big]
\label{vd}
\end{eqnarray}
where $f_i$s are given in Appendix. Note in the case of CMB radiation, $I$ can be total intensity of CMB or CMB thermal anisotropy (depending of angular dependence of $f_i$s)  while the contribution of $Q, U$ and $V$ are about or less than  $\%10$ of total CMB thermal anisotropy. As a result, to consider dominated contribution in our calculations Eqs.(\ref{qd}-\ref{vd}), we neglect terms in second order of  $Q, U$ and $V$. As  Eqs.(\ref{qd}-\ref{vd}) show, the initial circular polarization of an ensemble of photon $V(\mathbf{k})$ can be converted to linear one $U(\mathbf{k}),Q(\mathbf{k})$ and inverse due to Euler-Hisenberg interactions. To go further and
calculate  angular integrals most conveniently, we introduce the momentum and polarization vectors of incoming photons as follow \cite{cosowsky1994}
\begin{eqnarray}\label{unitvectors}
&&\hat{\mathbf{k}}=(\sin{\theta}\cos{\phi},\sin{\theta}\sin{\phi},\cos{\theta}),\nonumber\\
&&\vec{\hat{\epsilon}}_1(\mathbf{k})=(\cos{\theta}\cos{\phi},\cos{\theta}\sin{\phi},-\sin{\theta}),\\
&&\vec{\hat{\epsilon}}_2(\mathbf{k})=(-\sin{\phi},\cos{\phi},0).\nonumber
\end{eqnarray}
The exactly same definition are correct for momentum and  polarization vectors of target photons (denoted by $\mathbf{p}$ and $\vec{\epsilon}_s(\mathbf{p})$) just with $\theta\rightarrow\theta'$ and $\phi\rightarrow\phi'$. The angular integrals in Eqs.(\ref{qd}-\ref{vd}) must be done over $\theta'$ and $\phi'$. As momentum and polarization vectors of photons are defined in spherical coordinate, one can expand
all variables and Stokes parameters in terms of spherical harmonics $Y_l^m$ to make angular integrals easily, so we have
\begin{eqnarray}\label{spherical}
&&I(\mathbf{p})=\sum_{l'm'}I_{l'm'}(\mathbf{ p})Y_{l'}^{m'}(\theta',\phi'),\nonumber\\
&&(Q\pm i U)(\mathbf{p})=\sum_{l'm'}(Q\pm i U)_{l'm'}(\mathbf{p})Y_{l'}^{m'}(\theta',\phi'),\\
%&&(Q-iU)(\mathbf{\hat{p}})=\sum_{l'm'}a_{-2l'm'-2}Y_{l'}^{m'}(\theta',\phi')\nonumber\\
&&V(\mathbf{p})=\sum_{l'm'}V_{l'm'}(\mathbf{p})Y_{l'}^{m'}(\theta',\phi').\nonumber
\end{eqnarray}
Also we can use above equations to expand  $I(\mathbf{k})$, $Q(\mathbf{k})$, $U(\mathbf{k})$ and $V(\mathbf{k})$ in terms of spherical harmonics by replacing  $\theta\rightarrow\theta'$, $\phi\rightarrow\phi'$, $l'\rightarrow l$ and $m'\rightarrow m$.
So by considering the time evolution of Stokes parameters given in Eqs.(\ref{qd}-\ref{vd}), using expansions in  Eq.(\ref{spherical}) and adding the Compton scattering contributions to Euler-Heisenberg contributions, we have
\begin{eqnarray}\label{cegamma}
&&\frac{dI}{dt}=C^I_{e\gamma},\nonumber\\
&&\frac{d}{dt}(Q\pm i U)=C^{\pm}_{e\gamma}\mp i \dot{\kappa}_{\pm}V,\\
&&\frac{dV}{dt}=C^V_{e\gamma}+\dot{\kappa}_{U}U,\nonumber
\end{eqnarray}
where $C^I_{e\gamma}$, $C^{\pm}_{e\gamma}$ and $C^V_{e\gamma}$ denote contributions of Compton scattering  which their expressions could be found in \cite{cosowsky1994,zalda,hu}.
The Euler-Heisenberg contribution coefficients are given as following
\begin{eqnarray}
&&\dot{\kappa}_\pm=\frac{8\alpha^2}{45m^4 k^0}\!\!\!\int\!\!\!\frac{p^2dpd\Omega'}{(2\pi)^3 2p^0}(p^0k^0)^2
\Big[(-2iU(\mathbf{p})\pm I(\mathbf{p}))f_1(\hat{p},\hat{k})\Big]\label{kdotpm}\\
&&\dot{\kappa}_U=\frac{8\alpha^2}{45m^4 k^0}\!\!\! \int\frac{p^2dpd\Omega'}{(2\pi)^32p^0}(p^0k^0)^2
\Big[f_2(\hat{p},\hat{k})I(\mathbf{p})
\Big].\label{ku}
\end{eqnarray}
As shown in Eq.(\ref{kdotpm}), $\dot{\kappa}_\pm$ is divided to two term which are  proportional to  $U(p)$ and $I(p)$.  According to the earlier mentioned argument, to consider dominant contributions of Euler-Heisenberg effective Lagrangian in CMB power spectrum, we can neglect the term  including $U(p)$. Then
\begin{eqnarray}
\dot{\kappa}_\pm &=& \pm\frac{1}{15\pi}\,\, \sigma_T\, \frac{k}{m_e}\, \frac{I_0}{m_e} \Big(\int\,\frac{d^3p}{(2\pi)^3}p\, f_1(\hat{p},\hat{k}) \sum_{lm}Y_{l,m}\frac{I_{lm}(\mathbf{p})}{I_0}\Big),\nonumber\\
&=& \pm\dot{\tilde{\kappa}} \Big(f_1^0+\int\,\frac{d^3p}{(2\pi)^3}p\, \tilde{f}_1(\hat{p},\hat{k}) \sum_{lm}Y_{l,m}\frac{I_{lm}(\mathbf{p})}{I_0}\Big)\label{kappapm}\\
\dot{\kappa}_U &=& \frac{1}{15\pi}\,\, \sigma_T\, \frac{k}{m_e}\, \frac{I_0}{m_e} \Big(\int\,\frac{d^3p}{(2\pi)^3}p\, f_2(\hat{p},\hat{k}) \sum_{lm}Y_{l,m}\frac{I_{lm}(\mathbf{p})}{I_0}\Big),\nonumber\\
&=& \dot{\tilde{\kappa}} \Big(f_2^0+\int\,\frac{d^3p}{(2\pi)^3}p\, \tilde{f}_2(\hat{p},\hat{k}) \sum_{lm}Y_{l,m}\frac{I_{lm}(\mathbf{p})}{I_0}\Big),\label{kappau}
\end{eqnarray}
where $\dot{\tilde{\kappa}}=\frac{1}{15\pi}\,\, \sigma_T\, \frac{k}{m_e}\, \frac{I_0}{m_e}$  and here we separate $f_i(\hat{p},\hat{k})= f_{i}^0+\tilde{f}_i(\hat{p},\hat{k})$, note $f_{i}^0$ is constant part of $f_i(\hat{p},\hat{k})$ and also
\begin{eqnarray}
\int\frac{p\, d^3p}{(2\pi)^3} I(\mathbf{p}) = I_0(\bar{p})\simeq \bar{p}~ n_{\gamma}.
\end{eqnarray}
and $\bar {p}=|\mathbf{p}|$ is the average value of the momentum of target (CMB-photons). Be ware in above equations, the term including $\tilde{f}_i(\hat{p},\hat{k})$ is in the order of CMB temperature anisotropy $\sim\frac{\delta T}{T}$ which several order of magnitude smaller than the term including $f_i^0$. So it is reasonable to ignore the term including $\tilde{f}_i(\hat{p},\hat{k})$ for the rest of our calculation. As a result, by considering non-linear photon-photon interaction, a linear polarization converts to circular one while crossing through an isotopic unpolarized medium beam $I_0$.\\
To understand the above results,  we can assume that linearly polarized CMB photons encounter by an isotopic background magnetic and electric fields when they cross through the unpolarized beam. By purposing the mentioned point, we can rewrite Euler-Heisenberg  Hamiltonian by replacing   $F_{\mu\nu}\rightarrow B_{\mu\nu}+F_{\mu\nu}$ where $B_{\mu\nu}$ indicates background fields [for example see \cite{sorush}]
\begin{eqnarray}
H^0_I(t) &=& -\frac{\alpha^2}{90m^4_e}\int d^3x\Big(
[(F_{\mu\nu}+B_{\mu\nu})(F^{\mu\nu}+B^{\mu\nu}) ]^2+\frac{7}{4}[(F_{\mu\nu}+B_{\mu\nu})(\tilde{F}^{\mu\nu}+\tilde{B}^{\mu\nu})]^2\Big).
\label{int1}
\end{eqnarray}
Note in above equation, we just need terms with two $F_{\mu\nu}$ while terms including $(B_{\mu\nu}B^{\mu\nu})(F_{\mu\nu}F^{\mu\nu})$ and $(B_{\mu\nu}\tilde{B}^{\mu\nu})(F_{\mu\nu}\tilde{F}^{\mu\nu})$ do not affect in our results. So by using Eq.(\ref{A00}),
\begin{align}\label{int2}
H^0_I(t)=\frac{4\alpha^{2}}{90m^{4}_e}\int\frac{d^{3}p}{(2\pi)^{3}(2p^{0})^{2}}\sum_{ss^{'}}\hat a^{\dagger}_{s}(p)\hat a_{s^{'}}(p)[p^{\mu}B_{\mu\nu}\epsilon_{s^{'}}^{\nu}p^{\lambda}B_{\lambda\rho}\epsilon^{\ast\rho }_{s}-\frac{7}{4}
p^{\mu}\tilde{B}_{\mu\nu}\epsilon_{s^{'}}^{\nu}p^{\lambda}\tilde{B}_{\lambda\rho}\epsilon^{\ast\rho }_{s}].
\end{align}
and by substituting below equations
 \begin{eqnarray}
  p^{\mu}B_{\mu\nu}\epsilon_{s}^{\nu} &=& \vec{B}.(\vec{p}\times\vec{\epsilon}_{s})+p^0\vec{E}.\vec{\epsilon}_{s},\nonumber\\
  p^{\mu}\tilde{B}_{\mu\nu}\epsilon_{s^{'}}^{\nu} &=&2 \vec{E}.(\vec{p}\times\vec{\epsilon}_{s})+2p^0\vec{B}.\vec{\epsilon}_{s},\label{int3}
\end{eqnarray}
in Eq.(\ref{int2}), we obtain
\begin{align}\label{int4}
H^0_I(t)&&=\frac{4\alpha^{2}}{90m^{4}_e}\int\frac{d^{3}p}{(2\pi)^{3}(2p^{0})^{2}}\sum_{ss^{'}}\hat a^{\dagger}_{s}(p)\hat a_{s^{'}}(p)
\Big([(\vec{B}.(\vec{p}\times\vec{\epsilon}_{s})+p^0\vec{E}.\vec{\epsilon}_{s})(\vec{B}.(\vec{p}\times\vec{\epsilon}_{s'})+p^0\vec{E}.\vec{\epsilon}_{s'})]\nonumber\\
&&-7[
(\vec{E}.(\vec{p}\times\vec{\epsilon}_{s})+p^0\vec{B}.\vec{\epsilon}_{s})(\vec{E}.(\vec{p}\times\vec{\epsilon}_{s'})+p^0\vec{B}.\vec{\epsilon}_{s'})]\Big).
\end{align}
At the end, we have used Eqs.(\ref{bo}) and (\ref{int4}) to obtain the time evolution of Stokes parameters, here we just discuss $V$-parameter
\begin{align}\label{vdot1}
\dot V(\vec{k})=\frac{4\alpha^{2}k^{0}}{90m^{4}_e} \Big[\tilde{g}\,Q(\vec{k})+\tilde{f}\,U(\vec{k})\Big]
\end{align}
where
\begin{eqnarray}
  \tilde{g} &=& 2\Big(\vec B \cdot(\hat k\times\hat \epsilon_{2})\ \vec B.(\hat k\times\hat\epsilon_{1^{}})+\vec E\cdot\hat \epsilon_{2}\vec B\cdot(\hat k\times\hat\epsilon_{1^{}})+\vec E\cdot\hat \epsilon _{1} \vec B\cdot(\hat k\times\hat\epsilon_{2^{}})+\vec E\cdot\hat \epsilon _{1} \vec E\cdot\hat \epsilon_{2}\Big) \nonumber \\
  &+& 14\Big(\vec E \cdot(\hat k\times\hat \epsilon_{2})\ \vec E.(\hat k\times\hat\epsilon_{1})+\vec B\cdot\hat \epsilon_{2}\vec E\cdot(\hat k\times\hat\epsilon_{1})+\vec B\cdot\hat \epsilon_{1} \vec E\cdot(\hat k\times\hat\epsilon_{2})+\vec B\cdot\hat \epsilon _{1} \vec B\cdot\hat \epsilon_{2}\Big)\label{VQ}
\end{eqnarray}
and
\begin{equation}\label{VU}
    \tilde{f}=k^0\Big[6\big((\vec B\cdot\epsilon_{1})^2-(\vec B\cdot\epsilon_{2})^2)+6\big((\vec E\cdot\epsilon_{2})^2-(\vec E\cdot\epsilon_{1})^2)+16\big((\vec E\cdot\epsilon_{1})(\vec B\cdot\epsilon_{2})-(\vec B\cdot\epsilon_{1})(\vec E\cdot\epsilon_{2}))\Big].
\end{equation}
Now we are ready to check the results discussed in Eq.(\ref{kappau}).  Using Eqs. (\ref{unitvectors}) and considering a random direction for electric fields $\vec{E}=E(\sin{\theta_E}\cos{\phi_E},\sin{\theta_E}\sin{\phi_E},\cos{\theta_E})$, we will rewrite the average value of $<\tilde{f}>$ and $<\tilde{g}>$ as following
\begin{equation}\label{VU1}
    <\tilde{f}>=3/4(1-\cos{2\theta})<E^2>+<\tilde{f}_1(\theta_E,\phi_E)>\propto3/4(1-\cos{2\theta})I_0+<\tilde{f}_1(\theta_E,\phi_E)>
\end{equation}
note in above equation $3/4(1-\cos{2\theta})<E^2>$ is independent from the direction of electric fields as well as the polarizations of radiation ($<E^2>\propto I_0$). But $<\tilde{g}>$ does not include a term which can be independent from the direction of electric fields. In the simple word, $<\tilde{f}>$ has a contribution from isotropic unpolarized CMB radiation which comes from the nature of non-linear interaction between CMB photons themselves via Euler-Heisenberg Hamiltonian.

\section{The Time Evolution of CMB Polarizations due to Euler-Heisenberg Lagrangian and Compton Scattering}
In present section, we consider our rest calculation in the presence of the primordial scalar perturbations indicating by $(S)$ which we expand these perturbations in the Fourier modes characterized by a wave number $\mathbf{K}$. For each given
wave number $\mathbf{K}$, it is useful  to select a coordinate system
with $\mathbf{K} \parallel \hat{\mathbf{z}}$ and
$(\hat{\mathbf{e}}_1,\hat{\mathbf{e}}_2)=(\hat{\mathbf{e}}_\theta,
\hat{\mathbf{e}}_\phi)$.  
Temperature anisotropy $\Delta^{(S)}_{I}$, linear
polarizations ($\Delta^{(S)}_{Q}$ and $\Delta^{(S)}_{U}$) and circular polarization $\Delta^{(S)}_{V}$ of the CMB radiation can be expanded in an
appropriate spin-weighted basis as following \cite{zalda}
\begin{eqnarray}
&&
\Delta^{(S)}_{I}(\mathbf{K},\mathbf{k},\tau)=\sum_{\ell m}a_{\ell m}(\tau,K)Y_{lm}(\mathbf{n}),\label{AA0}\\
&&
\Delta^{\pm (S)}_{P}(\mathbf{K},\mathbf{k},\tau)=\sum_{\ell m}a_{\pm2,\ell m}(\tau,K) _{\pm2}Y_{lm}(\mathbf{n}),\label{AA}\\
&&
\Delta^{(S)}_{V}(\mathbf{K},\mathbf{k},\tau)=\sum_{\ell m}a_{V,\ell m}(\tau,K)Y_{lm}(\mathbf{n}),\label{AAV}
\end{eqnarray}
where we define
\bea
\Delta^{(S)}_I(\mathbf{K},\mathbf{k},\tau)=\left(4k\frac{\partial I_0}{\partial k}\right)^{-1} \Delta^{(S)}_ I(\mathbf{K},\mathbf{k},t),\,\,\,\,\,\,\,\Delta _{P}^{\pm (S)}=\left(4k\frac{\partial I_0}{\partial k}\right)^{-1}(Q^{(S)}\pm iU^{(S)}).
\eea
As usual, one can transfer the CMB temperature and polarizations $\Delta_{I,P,V}(\eta,\mathbf{K},\mu)$ in the conformal time $\eta$ and describe them by multi-pole moments as following
\begin{eqnarray}
\Delta_{I,P,V}(\eta,\mathbf{K},\mu)=\sum^{\infty}_{l=0}(2 l+1)(-i)^l\Delta^l_{I,P,V}(\eta,\mathbf{K})P_{l}(\mu)
\end{eqnarray}
where $\mu = \hat{n}\cdot\hat{\mathbf{K}} = \cos \theta$, the $\theta$ is angle between the CMB photon direction $\hat{n} = \mathbf{k}/|\mathbf{k}|$ and the wave vectors
$\mathbf{K}$, and $P_l(\mu)$ is the Legendre polynomial of rank $l$.
Here we should define left hand sides of Eq.(\ref{cegamma}) $\frac{d}{dt}$ to take into account  space-time structure and gravitational effects such as red-shift and so on. For each plane wave, each scattering and interaction can be described as the transport through a plane parallel medium \cite{mukh,chandra}, and finally  Boltzmann equations in the presence of the primordial scalar perturbations are given as
\begin{eqnarray}
&&\frac{d}{d\eta}\Delta_I^{(S)} +iK\mu \Delta_I^{(S)}+4[\dot{\psi}-iK\mu \varphi]
=\dot\tau_{e\gamma}\Big[-\Delta_I^{(S)} +
\Delta_{I}^{0(S)} +i\mu v_b +{1\over 2}P_2(\mu)\,\Pi\Big]  \label{Boltzmann}\\
&&\frac{d}{d\eta}\Delta _{P}^{\pm (S)} +iK\mu \Delta _{P}^{\pm (S)} = \dot\tau_{e\gamma}\Big[
-\Delta _{P}^{\pm (S)} -{1\over 2} [1-P_2(\mu)]\, \Pi\Big]\mp i\,a(\eta) \,\dot{\tilde{\kappa}}\,f_1^0\,\Delta _{V}^{(S)}
\label{Boltzmann1}\\
&&\frac{d}{d\eta}\Delta _{V}^{(S)} +iK\mu \Delta _{V}^{(S)} = -\dot\tau_{e\gamma}\Big[\Delta _{V}^{(S)}-\frac{3}{2}\mu \Delta _{V1}^{(S)}\Big]+\frac{i}{2}\dot{\tilde{\kappa}}\,f_2^0\,(\Delta^{-(S)}_P-\Delta^{+(S)}_P)
\label{Boltzmann2}
\end{eqnarray}
where $\dot{\tau}_{e\gamma}\equiv \frac{d\tau_{e\gamma}}{d\eta}$ which $\tau_{e\gamma}$ is Compton scattering optical depth, $a(\eta)$ is normalized scale factor and $\Pi\equiv \Delta_I^{2(S)}+\Delta_P^{2(S)}+\Delta _P^{0(S)}$.

The values of
$\Delta _{P}^{\pm (S)}(\hat{n})$ and $\Delta _{V}^{(S)}(\hat{n})$  at
the present time $\eta_0$ and the direction $\hat{n}$ can be obtained in following general form  by integrating of the Boltzmann equation (Eq's. (\ref{Boltzmann}-\ref{Boltzmann2})) along the line of sight \cite{zalda} and summing over all the Fourier modes $\mathbf{K}$ as follows
\begin{eqnarray}
\Delta _{P}^{\pm (S)}(\hat{\bf{n}})
&=&\int d^3 \bf{K} \xi(\bf{K})e^{\pm2i\phi_{K,n}}\Delta _{P}^{\pm (S)}
(\mathbf{K},\mathbf{k},\eta_0),\,\,\,\,\,\label{Boltzmann03}\\
\Delta _{V}^{ (S)}(\hat{\bf{n}})
&=&\int d^3 \bf{K} \xi(\bf{K})\Delta _{V}^{(S)}
(\mathbf{K},\mathbf{k},\eta_0),\,\,\,\,\,\label{Boltzmann3}
\end{eqnarray}
where $\phi_{K,n}$ is the angle needed to
rotate the $\mathbf{K}$ and $\hat{\bf{n}}$ dependent basis to a fixed
frame in the sky, $\xi(\mathbf{K})$ is a random variable using to
characterize the initial amplitude of each primordial scalar perturbations mode, and also the values of
$\Delta _{P}^{\pm (S)}(\mathbf{K},\mathbf{k},\eta_0)$ and $\Delta _{V}^{(S)}(\mathbf{K},\mathbf{k},\eta_0)$  are given as
\bea
 \Delta _{P}^{\pm (S)}
(\mathbf{K},\mu,\eta_0)&=&\int_0^{\eta_0} d\eta\,\dot\tau_{e\gamma}\,
e^{ix \mu -\tau_{e\gamma}}\,\,\Big[ {3 \over 4}(1-\mu^2)\Pi(K,\eta)\nonumber\\
&\mp&i\,f_1^0\,\frac{\dot{\tilde{\kappa}}}{\dot\tau_{e\gamma}}\,\Delta _{V}^{(S)}\Big],\label{EBS}
\end{eqnarray}
and
\bea
 \Delta _{V}^{(S)}
(\mathbf{K},\mu,\eta_0)
&\approx&\int_0^{\eta_0} d\eta\,
\dot\tau_{e\gamma}\,e^{ix \mu -\tau_{e\gamma}}\,\,\Big[ \frac{3}{2}\mu\Delta _{V1}^{(S)}-i\,f_2^0\,\frac{\dot{\tilde{\kappa}}}{\dot\tau_{e\gamma}}\,\,\Delta _{P}^{(S)}\Big],\label{VS}
\eea
in which $x=K(\eta_0 - \eta)$, $f_{1,2}({\hat{p},\hat{k}})$ are defined in (\ref{f1},\ref{f2}) and
\begin{equation}
   \Delta _{P}^{(S)}
(\bf{K},\mu,\eta)=\int_0^{\eta} d\eta\,\dot\tau_{e\gamma}\,
e^{ix \mu -\tau_{e\gamma}}\,\,\Big[ {3 \over 4}(1-\mu^2)\Pi(K,\eta)\Big].\label{PS}
\end{equation}
  The differential optical depth $\dot\tau_{e\gamma}(\eta)$ and total optical depth $\tau_{e\gamma}(\eta)$ due to the Thomson scattering at time  $\eta$ are defined as
\begin{equation}\label{optical}
    \dot{\tau}_{e\gamma}=a\,n_e\,\sigma_T,\,\,\,\,\,\,\,\tau_{e\gamma}(\eta)=\int_\eta^{\eta_0}\dot{\tau}_{e\gamma}(\eta) d\eta.
\end{equation}
%As is shown in (\ref{Boltzmann}), the temperature  anisotropy
%$\Delta_I^{(S)}$ doesn't have any source due to the forward
%Compton scattering  in the NC space-time therefore, we only focus on the other equations to explore the NC effects.  Meanwhile, (\ref{Boltzmann03}) and
%(\ref{Boltzmann3}) indicate that the effect of non-commutativity on the linear and circular polarization can be valuable for a significant value of $\frac{\kappa_{NC}}{\dot\tau_{e\gamma}}$ which is defined as follows
%\begin{equation}\label{eq1}
  %\tilde{\kappa}=  \frac{\kappa_{NC}}{\dot\tau_{e\gamma}}=\frac{3}{4}\frac{1}{\alpha}\,\frac{m_e^2}{\Lambda^2}\sum_{f=e,p}\frac{m_f}{k^0},
%\end{equation}
%which leads to larger values for Protons than the electrons.
%%%%%%%%%%%%%%%%%%%%%%%%%%%%%%%%%%%%%%%%%%%%%%%%%%%%%%%%%%%%%%%%%%%%%%%%%%%%%%%%%%%%%%%%%%%%%
\section{ The contribution of Euler-Heisenberg interaction for the circular power spectrum of CMB}
%%%%%%%%%%%%%%%%%%%%%%%%%%%%%%%%%%%%%%%%%%%%%%%%%%%%%%%%%%%%%%%%%%%%%%%%%%%%%%%%%%%%%%%%%%%%%
In the preceding  section, we have prepared all instruments  to calculate different power spectra $C_{l}^{X(S)}$s of CMB radiation due to Compton scattering and photon-photon forward scattering via Euler-Heisenberg interaction. So the power spectrum $C_{l}^{X(S)}$ in the presence of primordial scalar perturbation (indicating by $(S)$) is given as
\begin{equation}\label{PS1}
    C_{l}^{X(S)}=\frac{1}{2l+1}\sum_m\Big<a^*_{X,lm}\,a_{X,lm}\Big>,\,\,\,\,\,\, X=\{I,E,B,V\},
\end{equation}
where
\begin{eqnarray}
  a_{E,lm} &=& -(a_{2,lm}+a_{-2,lm})/2 ,\label{ae}\\
  a_{B,lm} &=& i(a_{2,lm}-a_{-2,lm})/2 ,\label{ab} \\
  a_{V,lm} &=& \int\,d\Omega Y^*_{lm} \Delta_V. \label{av}
\end{eqnarray}
By using (\ref{EBS}-\ref{PS}), the circular power spectrum $C^{V(S)}_{l}$ of CMB radiation can be written as following
 \begin{eqnarray}
 & &  C^{V(S)}_{l}=\frac{1}{2l+1}\sum_m\Big<a^*_{V,lm}\,a_{V,lm}\Big>,\nonumber \\
  &\approx&\frac{1}{2l+1}\int d^3{\bf K} P_{\phi}^{(S)}({\bf K},\eta)\sum_m\Big|\int d\Omega Y^*_{lm}\int_0^{\eta_0} d\eta\,
\dot\tau_{e\gamma}\,e^{ix \mu -\tau_{e\gamma}}\,\,\eta_{EH}(\eta)\,\,\Delta _{P}^{(S)}\Big|^2,~~\label{CVl}
\end{eqnarray}
where $\eta_{EH}(\tau)=f_2^0\,\frac{\dot{\tilde{\kappa}}}{\dot\tau_{e\gamma}}$
\bea
& & P_{\phi}^{(S)}({\bf K},\tau)\delta({\bf K'}-{\bf K})=\Big<\xi({\bf K})\,\xi({\bf K'})\Big>,
\eea
and $P_{\phi}^{(S)}({\bf K},\tau)$ is  the scalar power spectrum of primordial matter perturbations.

\begin{figure}
  \includegraphics[width=4in]{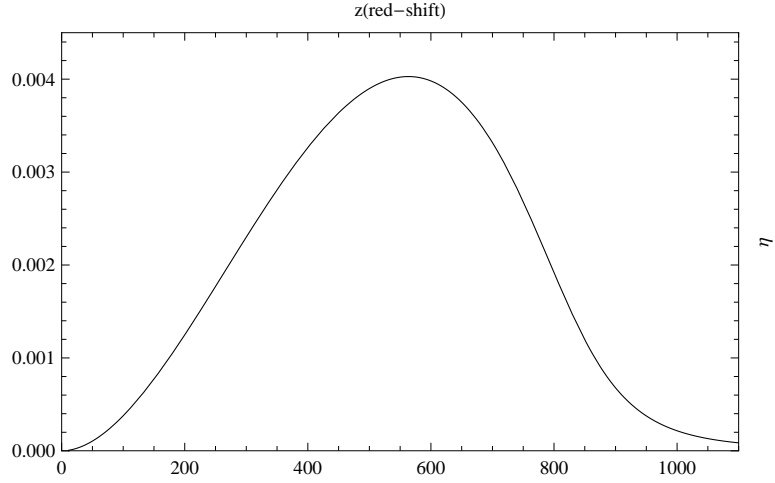}\\
  \caption{$\eta_{EH}(z)$ is plotted in terms of red-shift.}\label{etaz}
\end{figure}

Furthermore as shown Eq.(\ref{CVl}), the circular polarization cannot be generated in the scalar perturbation without considering the effects of Euler-Hiesenberg interactions. This result is in agreement with results of standard cosmology models \cite{cosowsky1994}. With this knowledge that $\dot{\tilde{\kappa}}$ and $\dot\tau_{e\gamma}$ depend on red-shift, we have
\begin{equation}\label{eta}
   \eta_{EH}(z)\,\simeq \,\dfrac{f_2^0}{15\pi}\dfrac{n_\gamma^0}{n_e^0}\frac{(1+z)^2}{\chi_e(z)}(\dfrac{T^0_{CMB}}{m_e})^2,
\end{equation}
 where $\chi_e(z)$ is fraction of free cosmic electron, $n_\gamma^0$ and $n_e^0$ are number densities of CMB photons and cosmic electrons at present time and $T^0_{CMB}\simeq 2.7K$. $\eta_{EH}(z)$ is plotted in terms of red-shift in Fig.(\ref{etaz}).

\begin{figure}
  \includegraphics[width=4in]{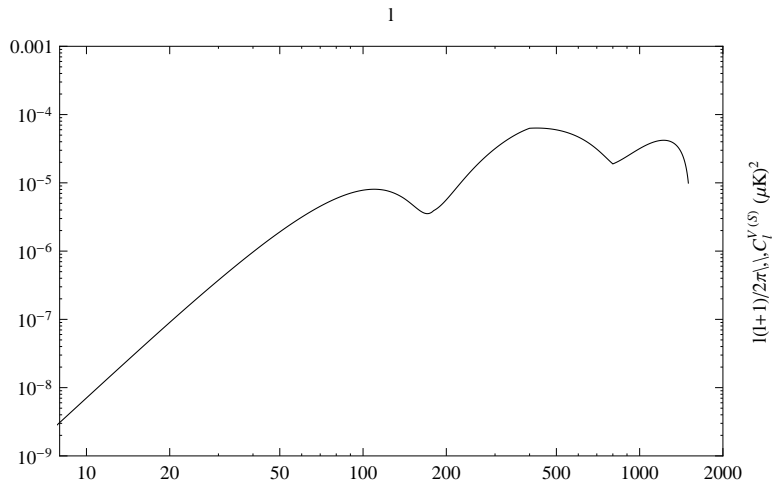}\\
  \caption{The power spectrum of circular polarization $l(l+1)/2\pi~~C_{l}^{V(S)}$ is plotted in terms of $l$ and in unit $(\mu K)^2$ due to Compton scattering and photon-photon forward scattering via Euler-Heisenberg Effective Lagrangian. This file contains the LCDM power spectra that are derived from Planck (2015) parameters and also we have modified CMBquick mathematica code to make above plot.}\label{ClVS}
\end{figure}

Now we can estimate $C^{V(S)}_{l}$ in terms of the linearly polarized power spectrum $C^{P(S)}_{l}$ and the average value of $\eta_{EH}$ as
\begin{equation}\label{CVl1}
   C_{l}^{V(S)}\, \approx\,(\eta_{EH}^{av})^2\, C^{P(S)}_{l},
\end{equation}
where
 \begin{eqnarray}
  C^{P(S)}_{l}
  =\frac{1}{2l+1}\int d^3{\bf K} P_{\phi}^{(S)}({\bf K},\tau)\sum_m\Big|\int d\Omega Y^*_{lm}\int_0^{\eta_0} d\eta\,
\dot\tau_{e\gamma}\,e^{ix \mu -\tau_{e\gamma}}\,\,\Delta _{P}^{(S)}\Big|^2,~~\label{Cpl}
\end{eqnarray}
and
\begin{eqnarray}
   \eta_{EH}^{av}&=&\frac{1}{z^{lss}}\int_0^{z^{lss}}\,\eta_{EH}(z)\,dz\simeq 0.0002,\label{eta-av}
\end{eqnarray}
where $z^{lss}$ indicates red-shift at last scattering surface.
 Using  the experimental value for $C^{P(S)}_{l}$ which is in the order of $\sim \mu K^2$  and Eqs.(\ref{CVl1})-(\ref{eta-av}), one can obtain  an estimation on the range of $C_{l}^{V(S)}\sim 10nK^2$, which is in the range of future experimental values. Note,  we just make  above estimation to have a sense about the contribution of Euler-Heisenberg interactions for the power spectrum of CMB circular polarization.  The more precisely estimation of $l(l+1)\,C_l^{V(S)}/(2\pi)$ is given in Fig.(\ref{ClVS}). Let's compare our results with experimental data reported by SPIDER group \cite{spider}. Constrains of the circular power spectrum $l(l+1)\,C_l^V/(2\pi)$ reported by SPIDER group is  in ranging from  141 to 203 $\mu K^2$ at 150 GHz for a thermal CMB spectrum and $33<l<307$ which is very larger than what can be found by considering non-linear photon-photon interaction. This means that if the results reported by \cite{spider} is confirmed, we have to search for another mechanisms (instead of non-linear CMB-CMB photons interaction) to satisfy them.

The Euler-Heisenberg interactions not only can generate circular polarization for CMB, but also generate the B-mode polarization in the presence of
scalar metric perturbations  in contrast with standard cosmology models \cite{zalda,zal}.
Next, one can divide the CMB
linear polarization  in terms of the divergence-free part (B-mode
$\Delta_{B}^{(S)}$) and the curl-free part (E-mode $\Delta_{E}^{(S)}$)
which are defined in terms of Stokes parameters as following
\begin{eqnarray}\label{Emode}
\Delta_{E}^{(S)}({\bf \hat{n}})&\equiv&-\frac{1}{2}[\bar{\eth}^{2}\Delta_{P}^{+(S)}(\hat{\bf{n}})+\eth^{2}\Delta_{P}^{-(S)}(\hat{\bf{n}})],\\
\label{Bmode}\Delta_{B}^{(S)}({\bf \hat{n}})&\equiv&\frac{i}{2}[\bar{\eth}^{2}\Delta_{P}^{+(S)}(\hat{\bf{n}})-\eth^{2}\Delta_{P}^{-(S)}(\hat{\bf{n}})],
\end{eqnarray}
where $\eth$ and $\bar{\eth}$ indicate  spin raising and lowering
operators respectively \cite{zal}. As Eqs.(\ref{Boltzmann3}),(\ref{PS1}),(\ref{eta-av}) and (\ref{Bmode}) shown, the  B-mode
 power spectrum $C^{B(S)}_{l}$  is given in terms of
the circular polarization power spectrum $C^{V(S)}_{l}$ which can be estimated as
\begin{equation}\label{Bmode1}
    C^{B(S)}_{l}\propto \bar{\eta}^2 C_{l}^{V(S)}\ll n K^2.
\end{equation}
 Note the B-mode generating by Euler-Hiesenberg interaction is very small than $n K^2$ and so that we can neglect it.
%%%%%%%%%%%%%%%%%%%%%%%%%%%%%%%%%%%%%%%%%%%%%%%%%%%%%%%%%%%%%%%%%%%%%%%%%%%%%%%%%%%%%%%%%%%%%%%%%%%%%%%%%%%%

\section{\large Conclusion and remarks}
In this work, we have solved
the first order of the Quantum
Boltzmann Equation for the density matrix of CMB radiation by considering Compton scattering and non-linear photon-photon forward scattering via the Euler-Heisenberg effective
Lagrangian as collision terms. We have shown
 that propagating photons convert their linear polarizations to circular
polarizations via  the Euler-Heisenberg effective
interaction. Also we have discussed that by considering non-linear CMB-CMB photons interaction, CMB linear polarization converts to circular one while crossing through CMB isotopic unpolarized medium $I_0$. The power spectrum of circular polarization in CMB radiations $ C^{V(S)}_{l}$ in the presence of scalar perturbations is given in terms of linearly polarized power spectrum of CMB radiation $C^{V(S)}_{l}\sim(\eta^{av}_{EH})^2 C^{P(S)}_{l}$ which $\eta_{EH}$ (\ref{eta}) is given in terms of redshift by factor $(1+z)^2/\chi_e(z)$ and also  $\eta^{av}_{EH}\simeq 0.0002$ (\ref{eta-av}).
Also,  we
have estimated the average  value of circular power spectrum is $1(l+1)\,C_{l}^{V(S)}/(2\pi)\sim 10^{-4}\mu\,K^2$ for $l\sim300$ at present time which is very smaller than recently reported data (SPIDER collaboration) but in the range of the future achievable experimental data. $l(l+1)\,C_l^{V(S)}/(2\pi)$ is plotted  in Fig.(\ref{ClVS}). As a result, it is necessary  to search for another mechanisms (instead of non-linear CMB-CMB photons interaction) to satisfy  SPIDER results for CMB circular polarization. 
We also show that the generation of B-mode polarization for CMB photons in the presence of the primordial scalar perturbation via Euler-Heisenberg interaction is possible however this contribution for B-mode polarization is not remarkable. It is shown in Eq.(\ref{Bmode1}) that $C^{B(S)}_{l}\ll nK^2$.

%\section*{\small Acknowledgment}
%I would like to thank R. Mohammadi and S. S. Xue  for fruitful discussion.
\newpage
\section{Appendix A}
The time-evolution of the density matrix approximately obtained as
\begin{eqnarray}
(2\pi)^3 \delta^3(0)2k^0
\frac{d}{dt}\rho_{ij}(k) \!\!&\approx& i\langle
\left[H^0_I(t),D^0_{ij}(k)\right]\rangle\nonumber\\
&=&-\frac{2\alpha^2i}{45m^4}(2\pi)^3\delta^3(0)\times\!\!\! \int\frac{d^3p}{(2\pi)^32p^0}\Bigg[(p.k)^2
[\epsilon_{s}(k).\epsilon_{s'}(p)\epsilon_{l}(k).\epsilon_{l'}(p)]\nonumber\\
&\times&\{-5\rho_{s'l'}(p)\rho_{is}(k)\delta^{lj}+5\rho_{s'l'}(p)\rho_{lj}(k)\delta^{si}
+4\rho_{l's'}(p)\rho_{lj}(k)
\delta^{si}
\nonumber\\&-&4\rho_{l's'}(p)\rho_{is}(k)\delta^{lj}
+3\rho_{l's'}(p)\rho_{sj}(k)\delta^{li}-3\rho_{l's'}(p)\rho_{il}(k)\delta^{sj}
\nonumber\\&+&4\rho_{s'l'}(p)\rho_{sj}(k)\delta^{li}-4\rho_{s'l'}(p)\rho_{il}(k)\delta^{sj}
+9\rho_{lj}(k)\delta^{si}\delta^{s'l'}\nonumber\\&-&9\rho_{is}(k)\delta^{lj}\delta^{s'l'}+3\rho_{sj}(k)\delta^{s'l'}\delta^{li}
-3\rho_{il}(k)\delta^{sj}\delta^{s'l'}\}
\nonumber\\&+&[p.\epsilon_{s}(k)k.\epsilon_{s'}(p)p.\epsilon_{l}(k)k.\epsilon_{l'}(p)
-2(p.k)\epsilon_{s}(k).\epsilon_{s'}(p)p.\epsilon_{l}(k)k.\epsilon_{l'}(p)]
\nonumber\\&\times&\{8\rho_{lj}(k)\delta^{si}\delta^{s'l'}-8\rho_{is}(k)\delta^{lj}\delta^{s'l'}
+4\rho_{l's'}(p)\rho_{lj}(k)\delta^{si}
\nonumber\\&-&4\rho_{l's'}(p)\rho_{is}(k)\delta^{lj}
-4\rho_{s'l'}(p)\rho_{is}(k)\delta^{lj}
+4\rho_{s'l'}(p)\rho_{lj}(k)\delta^{si}
\nonumber\\&+&4\rho_{sj}(k)\delta^{l's'}\delta^{li}+4\rho_{sj}(k)\rho_{l's'}(p)\delta^{li}-4\rho_{il}(k)\delta^{l's'}\delta^{sj}
\nonumber\\&-&4\rho_{l's'}(p)\rho_{il}(k)\delta^{sj}+4\rho_{s'l'}(p)\rho_{sj}(k)\delta^{li}
-4\rho_{s'l'}(p)\rho_{il}(k)\delta^{sj}\}\nonumber\\
&-&28\epsilon^{\mu\nu\alpha\beta}\epsilon^{\sigma\nu'\gamma\beta'}k_\gamma
k_\mu p_\alpha
p_\sigma\epsilon_{s'\beta}(p)\epsilon_{l\nu'}(p)\epsilon_{s\nu}(k)\epsilon_{l'\beta'}(k)\nonumber\\
&\times&\left[\rho_{l'j}(k)\delta^{si}-\rho_{is}(k)\delta^{l'j}+\rho_{sj}(k)\delta^{l'i}-\rho_{il'}(k)\delta^{sj}\right]\nonumber\\
&\times&[\rho_{ls'}(p)+\rho_{s'l}(p)+\delta^{s'l}]\Bigg],
\label{j1}
\end{eqnarray}
where $k$ and $p$ indicate the energy-momentum states of photons
and $\delta^3(0)$ will be cancelled in the final expression.
Here detail of abbreviated functions in Eqs.~(\ref{id}-\ref{vd}) are brought.
%\begin{eqnarray}
%f_1(\hat{p},\hat{k})&=&2\Bigg[(\hat{p}.\hat{k})^2\Big(\hat{\epsilon}_{1}(k).\hat{\epsilon}_{1}(p)\hat{\epsilon}_{2}(k).\hat{\epsilon}_{1}(p)
%+\hat{\epsilon}_{2}(k).\hat{\epsilon}_{2}(p)\hat{\epsilon}_{1}(k).\hat{\epsilon}_{2}(p)\Big)\nonumber\\
%&+&\hat{p}.\hat{\epsilon}_{1}(k)\hat{p}.\hat{\epsilon}_{2}(k)\Big((\hat{k}.\hat{\epsilon}_{1}(p))^2+(\hat{k}.\hat{\epsilon}_{2}(p))^2\Big)\nonumber\\
%&-&(\hat{k}.\hat{p})\Bigg(\Big(\hat{\epsilon}_{1}(k).\hat{\epsilon}_{1}(p)\hat{p}.\hat{\epsilon}_{2}(k)+\hat{\epsilon}_{2}(k).\hat{\epsilon}_{1}(p)\hat{p}.\hat{\epsilon}_{1}(k)\Big)\hat{k}.\hat{\epsilon}_{1}(p)\nonumber\\
%&+&\Big(\hat{\epsilon}_{2}(k).\hat{\epsilon}_{2}(p)\hat{p}.\hat{\epsilon}_{1}(k)+\hat{\epsilon}_{1}(k).\hat{\epsilon}_{2}(p)\hat{p}.\hat{\epsilon}_{2}(k)\Big)\hat{k}.\hat{\epsilon}_{2}(p)\Bigg)
%\Bigg]
%\end{eqnarray}
\begin{eqnarray}\label{f1}
f_1(\hat{p},\hat{k})&=&2\Bigg[
(\hat{p}.\hat{k})^2\Big((\hat{\epsilon}_{2}(k).\hat{\epsilon}_{1}(p))^2-(\hat{\epsilon}_{1}(k).\hat{\epsilon}_{1}(p))^2+(\hat{\epsilon}_{2}(k).\hat{\epsilon}_{2}(p))^2-(\hat{\epsilon}_{1}(k).\hat{\epsilon}_{2}(p))^2\Big)\nonumber\\
&+&\Big((\hat{p}.\hat{\epsilon}_{2}(k))^2-(\hat{p}.\hat{\epsilon}_{1}(k))^2\Big)\Big((\hat{k}.\hat{\epsilon}_{2}(p))^2+(\hat{k}.\hat{\epsilon}_{1}(p))^2\Big)
\nonumber\\
&+&2(\hat{k}.\hat{p})\Bigg(\Big(\hat{\epsilon}_{1}(k).\hat{\epsilon}_{1}(p)\hat{p}.\hat{\epsilon}_{1}(k)
-\hat{\epsilon}_{2}(k).\hat{\epsilon}_{1}(p)\hat{p}.\hat{\epsilon}_{2}(k)\Big)\hat{k}.\hat{\epsilon}_{1}(p)\nonumber\\
&+&\Big(\hat{\epsilon}_{1}(k).\hat{\epsilon}_{2}(p)\hat{p}.\hat{\epsilon}_{1}(k)
-\hat{\epsilon}_{2}(k).\hat{\epsilon}_{2}(p)\hat{p}.\hat{\epsilon}_{2}(k)\Big)\hat{k}.\hat{\epsilon}_{2}(p)\Bigg)
\Bigg]
\end{eqnarray}
%\begin{eqnarray}
%f_3(\hat{p},\hat{k})&=&2\Bigg[
%(\hat{p}.\hat{k})^2\Big((\hat{\epsilon}_{2}(k).\hat{\epsilon}_{1}(p))^2-(\hat{\epsilon}_{1}(k).\hat{\epsilon}_{1}(p))^2-(\hat{\epsilon}_{2}(k).\hat{\epsilon}_{2}(p))^2+(\hat{\epsilon}_{1}(k).\hat{\epsilon}_{2}(p))^2\Big)\nonumber\\
%&+&\Big((\hat{p}.\hat{\epsilon}_{2}(k))^2-(\hat{p}.\hat{\epsilon}_{1}(k))^2\Big)\Big((\hat{k}.\hat{\epsilon}_{2}(p))^2-(\hat{k}.\hat{\epsilon}_{1}(p))^2\Big)
%\nonumber\\
%&+&2(\hat{k}.\hat{p})\Bigg(\Big(\hat{\epsilon}_{1}(k).\hat{\epsilon}_{1}(p)\hat{p}.\hat{\epsilon}_{1}(k)
%-\hat{\epsilon}_{2}(k).\hat{\epsilon}_{1}(p)\hat{p}.\hat{\epsilon}_{2}(k)\Big)\hat{k}.\hat{\epsilon}_{1}(p)\nonumber\\
%&-&\Big(\hat{\epsilon}_{1}(k).\hat{\epsilon}_{2}(p)\hat{p}.\hat{\epsilon}_{1}(k)
%-\hat{\epsilon}_{2}(k).\hat{\epsilon}_{2}(p)\hat{p}.\hat{\epsilon}_{2}(k)\Big)\hat{k}.\hat{\epsilon}_{2}(p)\Bigg)
%\Bigg]
%\end{eqnarray}
\begin{eqnarray}
f_2(\hat{p},\hat{k})&=&2\Bigg[(\hat{p}.\hat{k})^2\Big([\hat{\epsilon}_{2}(k).\hat{\epsilon}_{1}(p)]^2-[\hat{\epsilon}_{1}(k).\hat{\epsilon}_{1}(p)]^2
+[\hat{\epsilon}_{2}(k).\hat{\epsilon}_{2}(p)]^2-[\hat{\epsilon}_{1}(k).\hat{\epsilon}_{2}(p)]^2\Big)\nonumber\\
&+&2(\hat{p}.\hat{k})\Bigg(\Big(\hat{\epsilon}_{2}(k).\hat{\epsilon}_{2}(p)\hat{p}.\hat{\epsilon}_{2}(k)
-\hat{\epsilon}_{1}(k).\hat{\epsilon}_{2}(p)\hat{p}.\hat{\epsilon}_{1}(k)\Big)\hat{k}.\hat{\epsilon}_{2}(p)-\hat{\epsilon}_{1}(k).\hat{\epsilon}_{1}(p)
\hat{p}.\hat{\epsilon}_{1}(k)\hat{k}.\hat{\epsilon}_{1}(p)\Bigg)\nonumber\\
&+&\Big((\hat{p}.\hat{\epsilon}_{1}(k))^2-(\hat{p}.\hat{\epsilon}_{2}(k))^2\Big)\Big((\hat{k}.\hat{\epsilon}_{1}(p))^2+(\hat{k}.\hat{\epsilon}_{2}(p))^2
\Big)\Bigg].\label{f2}
\end{eqnarray}
\newpage
%%%%%%%%%%%%%%%%%%%%%%%%%%%%%%%%%%%%%%%%%%%%%%%%%%%%%%%%%%%%%%%%%%%%%%%%%%%%%%%%%%

\end{document}